\documentstyle[prd,aps]{revtex}
\begin{document}

\draft

\title{Collapse of Axionic Domain Wall and Axion Emission}
\author{Michiyasu Nagasawa}
\address{Department of Physics, School of Science, The University of
  Tokyo, Tokyo 113, Japan}
\author{Masahiro Kawasaki}
\address{Institute for Cosmic Ray Research, The University of Tokyo,
  Tanashi 188, Japan}
\date{\today}

\maketitle

\begin{abstract}
    We examine the collapse of an axion domain wall bounded by an
    axionic string. It is found that the collapse proceeds quickly and
    axion domain walls disappear. However axions are emitted in the
    collapse and its energy density increases during radiation
    dominated era and contributes significantly to the present mass
    density of the universe. In particular the axion emitted from the
    wall can account for the dark matter in the universe for $F_a
    \gtrsim 10^{10}\mbox{GeV}$.
\end{abstract}


\section{Introduction}

The axion\cite{Peccei,Wilczek,Kim,DFS} is the Nambu-Goldstone boson
associated with the Peccei-Quinn symmetry breaking which was invented
as the most natural solution to the strong CP problem\cite{tHoot} of QCD.
The Peccei-Quinn symmetry breaking scale $F_a$ is stringently
constrained by the consideration of accelerator experiments, stellar
cooling and cosmology. The allowed range of $F_a/N$ ( axion window ) is
between $10^{10}\mbox{GeV}$ and $10^{12}\mbox{GeV}$, where the integer $N$
is the color anomaly of Peccei-Quinn symmetry. 

Since the Peccei-Quinn symmetry is a global U(1) symmetry,  global
strings (axionic strings) are produced during spontaneous symmetry
breaking. At this
stage the potential for the axion field is flat, i.e., the axion is
massless.  However the axion has a mass at QCD scale
through the instanton effect. 
The potential is written by
\begin{equation}
   V(A) = f_{\pi}^2m_{\pi}^2\left( 1 - \cos\frac{NA}{F_a}\right),
   \label{potential}
\end{equation}
where $f_{\pi}$ is the pion decay constant and $m_{\pi}$ is its mass.
Then the mass of the axion is given by $m_a \simeq
f_{\pi}m_{\pi}/{F_a/N}$. This potential(\ref{potential}) has $Z_n$
symmetry and takes its minimum at $A = 0, (f_a/N)\pi, 2(f_a/N)\pi,
\ldots 2f_a\pi$. $Z_n$ discrete symmetry is spontaneously broken to
produce two dimensional topological defects, i.e., axionic domain walls.
The property of the axionic domain wall is characterized by $N$.
Since $N$ domain walls stretch out from each axionic string, the network
of the string-wall systems is very complicated and survives long enough
to dominate the universe for $N > 1$ axionic domain wall \cite{RPS}.
Therefore the $N>1$ domain wall is not accepted cosmologically.

One might expect that the density of the domain wall can be diluted in
the inflationary universe. However, for the dilution mechanism to
work, both the reheating temperature and the expansion rate during
inflation should be smaller than $F_a$ \cite{Linde}, which is rather
unnatural requirement for many inflation models. Therefore the domain
wall problem in the axion model is serious one.

In the case of $N=1$, the domain wall is a disk bounded by the axionic
string. The wall with the string boundary is no longer stable and it
might collapse by the surface tension.  In fact, since, as seen later,
the surface tension is stronger than the tension due to the string for
the domain wall with size much greater than the width of the wall, the
dynamics of the wall is controlled by the wall tension. Therefore the
string that bounds the wall cannot prevent the wall from collapsing.
Typically the wall has size of horizon length ($\sim t_{QCD}$) when it
is formed.  Therefore the time scale for collapse is about $t_{QCD}
\sim 10^{-4}$ sec and the walls disappear quickly without overclosing
the universe and hence the domain wall seems harmless in the case of
$N=1$ axion model.  However it is expected that a number of axions are
produced when the axionic walls collapse. If produced axions are
non-relativistic or cold, their energy density decreases slowly
($\propto a^{-3}$ $a$: scale factor) compared with the radiation
density ($ \propto a^{-4}$). Hence the relative contribution of axions
to the total density of the universe increases with $a$ until the
universe becomes matter dominated.

In this paper we study the collapse of the axionic domain wall ($N=1$)
by the numerical integration of equations of motion for the axion
field and estimate the energy density of axion field after the
collapse.  It is found that the energy carried by the walls is
converted to the axions, which gives significant contribution to the
mass density of the present universe and might account for the dark
matter of the universe.

The axion emission is also expected from axionic string before
$t_{QCD}$\cite{Davis} or annihilation of axionic domain
walls\cite{Lyth}.  However, since it is shown later that parallel two
domain walls go through each other without annihilation, it is
unlikely that a large number of axions are produced by interaction of
two domain walls.  The massless axion can be produced by axionic
strings and they acquire mass after the QCD phase transition, which
might give significant contribution to the total density of the
present universe. However there are two independent quantitative
estimations by Davis\cite{Davis} and by Harari and
Sikivie\cite{Hagmann} and, unfortunately, they are quite different.
The contribution from axionic domain walls is expected to be at least
comparable to that from axionic strings if the estimation by Harari
and Sikivie is correct.

\section{Dynamics of Domain Wall}

The dynamics of the Peccei-Quinn scalar field $\phi$ is described by
the lagrangian:
\begin{eqnarray}
    {\cal L} & = & \partial^{\mu} \phi^{*} \partial_{\mu} \phi 
    + \frac{\lambda}{4} [|\phi|^2 - F_a^2]^2
    \nonumber\\
    & & = \partial^{\mu} \phi^{*} \partial_{\mu} \phi + V_s(|\phi|)
   \label{eq:lagra}
\end{eqnarray}
where $\lambda$ is the coupling constant. As the universe cools down,
Peccei-Quinn U(1) symmetry ($\phi \rightarrow e^{i\theta}\phi$) is
spontaneously broken and the scalar field has vacuum expectation value
$\langle | \phi | \rangle = F_a$.  After global U(1) symmetry is
broken, axionic strings are formed. From causality argument, about one
axionic string is produced within the horizon at $T\simeq F_a$. Since
the line energy density of the string is $4\pi F_a^2$, the string
density $\rho_{st}$ is about $4\pi F_a^2 t/t^3 \sim 4\pi
F_a^4(F_a/m_{pl})^2$ which is only $\sim 10^{-13} - 10^{-17}$ of total
density of the universe at the formation epoch.  Defining
$\phi = |\phi|\exp(iA/F_a)$, the lagrangian for axion field
$A$ is derived from eq.(\ref{eq:lagra}) as
\begin{eqnarray}
    {\cal L}_a & = & \partial^{\mu} A \partial_{\mu} A 
    + f_{\pi}^2m_{\pi}^2\left( 1 - \cos\frac{A}{F_a}\right) ,
    \nonumber\\
    & & = \partial^{\mu} A \partial_{\mu} A + V_w(A)
   \label{eq:alagra}
\end{eqnarray}
where the second term comes from QCD instanton effect which gives
axion mass $m_a \simeq f_{\pi}m_{\pi}/{F_a}$.  The potential $V_w(A)$
has a minimum at $A = 0, 2\pi F_a$ and domain walls are produced
between $A=0$ phase and $A = 2\pi F_a$ phase. More precisely, the
axion mass has temperature dependence and increase as 
\begin{equation}
    m_a(T) \simeq 0.1m_a(T=0) (\Lambda_{QCD}/T)^{3.7},
\end{equation}
where $\Lambda_{QCD}$ is the QCD scale $\sim 200\mbox{MeV}$. The domain
walls are produced when the axion mass becomes greater than the
expansion rate of the universe, i.e., $m_a(T_1) \simeq
\dot{a}(T_1)/a$.  The axionic domain wall has size of about horizon
length at the formation time $t_1(T_1)$ and has the axionic string on
its boundary. The surface tension of the axionic domain wall is
$\sigma \simeq 16 m_a F_a^2$\cite{Vilenkin}.  For wall with size $R$,
the ratio of surface energy to string energy is given by
\begin{equation}
    \frac{\mu R}{\sigma R^2} = \frac{4\pi F_a^2}{16m_a F_a^2 R} \simeq
    \frac{1}{m_a R}.
    \label{walltostring}
\end{equation}
Therefore the dynamics of the axionic wall with size greater than
$R^{*} \simeq 1/(m_a) \gg t_1$ is determined by the wall tension and
the string tension can be neglected.

The cosmological evolution of the wall is determined by the surface
tension and interaction of walls; the former makes the wall shrink and
the latter cuts the wall into small pieces. In both cases the wall
finally collapses by the surface tension.  After collapse the domain
walls disappear and the energy that the wall had is converted into
kinetic energy of axion fields.  If the energy of axion fields changes
like non-relativistic particles as the universe expands, the axion
energy density is 
\begin{equation}
    \rho_a(t) = \rho_{wall}(t_1)\left(\frac{a(t_1)}{a(t)}\right)^3,
    \label{axiondensity}
\end{equation}
where $\rho_a(t)$ and $\rho_{wall}$ are the energy density of the
axion field and the axionic wall, respectively.  For relativistic
axion, the axion density decreases as $\sim a^{-4}$ until axion
becomes non-relativistic. Then eq.(\ref{axiondensity}) is changed to
\begin{equation}
    \rho_a(t) = \rho_{wall}(t_1)\left(\frac{a(t_1)}{a(t)}\right)^3 
    \left(\frac{\langle E_a \rangle}{m_a}\right)^{-1},
    \label{axiondensity2}
\end{equation}
where $\langle E_a \rangle$ is the average energy of emitted axion at
$t_1$.  Assuming the mean distance and the mean radius of
walls are $\alpha t_1$ and $\beta t_1$, $\rho_{wall}$ is given by
\begin{equation}
    \rho_{wall}(t_1) = 
    \frac{\sigma \pi (\beta t_1)^2}{(\alpha t_1)^3}
    = 16\pi\alpha^{-3}\beta^2 m_a F_a^2
    \left(\frac{16\pi^3 {\cal N}}{45}\right)^{1/2}\frac{T_1}{m_{pl}}, 
    \label{walldensity}
\end{equation}
where ${\cal N}$ is the relativistic degrees of freedom at $t_1\simeq 1
\mbox{GeV}$. We expect that the numerical parameters $\alpha$ and $\beta$
are O(1) from causality, although the precise values should be determined
by the realistic simulation of cosmological formation of the wall-string
system, which is beyond the scope of the present paper.
 
Using the entropy density of the universe $s (= 2\pi^2{\cal N}T^3/45)$
and its conservation  ( $s a^3 =$ const.),
the number density of the axion  can be written as
\begin{equation}
    \frac{n_a}{s} = 380 \alpha^{-3}\beta^2 {\cal N}^{-1/2} 
    \left(\frac{F_a^2}{T_1 m_{pl}}\right)
    \left(\frac{\langle E_a \rangle}{m_a}\right)^{-1}.
    \label{axion-entropy}
\end{equation}
Then the present density of the axion is 
\begin{equation}
    \rho_a = 1.06 \times 10^6 \mbox{cm}^{-3} \alpha^{-3}\beta^2
    {\cal N}^{-1/2} m_a 
    \left(\frac{F_a^2}{T_1 m_{pl}}\right)
    \left(\frac{\langle E_a \rangle}{m_a}\right)^{-1}.
    \label{present-axion-density}
\end{equation}
Since ${\cal N}=289/4$, $T_1 \simeq 2\mbox{GeV} (F_a/10\mbox{GeV})^{-0.18}$
and $m_a = 6.2\times 10^{-4}\mbox{eV}(F_a/10^{10}\mbox{GeV})^{-1}$, 
\begin{equation}
    \rho_a = 3.2 \times 10^2 \mbox{eV} \mbox{cm}^{-3} \alpha^{-3}\beta^2
    \left(\frac{F_a}{10^{10}\mbox{GeV}}\right)^{1.18}
    \left(\frac{\langle E_a \rangle}{m_a}\right)^{-1}.
\end{equation}
The contribution of the axion to the density parameter $\Omega$ is given by
\begin{equation}
    \Omega_a h^2 = 0.030 \alpha^{-3}\beta^2
    \left(\frac{F_a}{10^{10}\mbox{GeV}}\right)^{1.18}
    \left(\frac{\langle E_a \rangle}{m_a}\right)^{-1},
    \label{omega-axion}
\end{equation}
where $h$ is the Hubble constant in units of $100\mbox{km}/\sec/\mbox{Mpc}$.
Therefore the axion density is large enough to account for the dark
matter in the universe unless the axion is ultra-relativistic when it
is emitted.  We estimate the $\langle E_a \rangle/m_a$ by numerical
simulation in the next section.

\section{Simulation of Collapse}
\label{sec:simulation}

In order to follow the motion of domain walls, we have solved the
evolution equation of the axion field numerically. When a wall piece
much smaller than the cosmological horizon is considered, the cosmic
expansion can be ignored.
Then the field equation under the Minkowski background is written as
\begin{equation}
    \frac{\partial^2 \phi}{\partial t^2}-\nabla^2\phi
    =-\frac{~\partial}{\partial \phi}(V_s+V_w)~,
    \label{eq:eveq}
\end{equation}
using the potentials in the equations (\ref{eq:lagra}) and
(\ref{eq:alagra}).  We have employed the staggered leapfrog method to
solve the differential equation. The model parameters are chosen such
that the width of the wall is equal to ten simulation meshes and the
vacuum energy of the string is a hundred times larger than that of the
wall. The variation of the numerical value in the latter condition
does not alter our conclusion since only the wall tension governs the
motion of walls as we have mentioned above.  The boundary condition is
periodic, which is useful to check the accuracy of our calculation
since the total energy in the simulation box is conserved.  The
solution of a static infinite planar wall under $|\phi|=v$ is used in
the initial configuration.
When the wall lies in $yz-$ plane, it is expressed as
\begin{equation}
    \frac{A(x)}{F_a}=\pi +2\sin^{-1}(\tanh m_a x) ~.
\end{equation}
Thus the scale of the inverse axion mass represents the characteristic
thickness of the axionic wall. The basic numerical technique is
the same one in the previous paper.
See the reference \cite{Naga} for more details.

First we have confirmed that approaching walls that face in parallel
each other pass through one another. Widrow showed this is true in the
case of a toy sine-Gordon potential\cite{Widrow}. We have reproduced
the passing phenomenon by numerical simulations using the potential
(\ref{eq:eveq}).  Fig.\ref{fig:pass} shows the result. The initial
condition is set so that the relative velocity is $0.05c$, where $c$
is light velocity and the separation between walls is 200 meshes.
Notice that the size of one mesh is equal to $F_a^{-1}$ in our
simulation and we take $m_a = 0.1 F_a$. As the time evolves, two
walls become close and go away without any crush. When the initial
relative velocity is much larger, for example, equal to $0.5c$, the
pair annihilation of walls occurs neither.  Therefore it is unlikely
that annihilation of domain walls occurs and produces axions
during the collision process. The possibility which is described in
the reference \cite{Lyth} should be unfavorable.

On the other hand, in the case of the encounter of a wall with a wall
edge, {i.e.}, a string cuts the wall and the process of disintegration
advances\cite{Naga}.  Hence by repeated intercommutations, large walls
are broken to small pieces. When their size becomes comparable with
the thickness scale of the wall, they collapse and radiate energy as
axions. In order to see how fast and effective the wall collapse is,
we have performed the simulation of the evolution of a small wall piece.
As a simple example we followed the time evolution of a disk wall
surrounded by a circular string in the previous work \cite{Naga}.
The result shows that the disk wall shrinks at the
velocity of light and the energy of the wall is converted to axions.
We can say that the wall collapse process is so rapid that the
radiation of gravitational waves is hardly expected.

In order to see the wall collapse with higher accuracy, we have
performed the simulations of a strip wall. Figs.\ref{fig:obi}(a)(b)(c)
demonstrate the two-dimensional distribution of $V_w$. The first one
shows the initial configuration in which there is one strip wall of
infinite height and 20 meshes length. As the time evolves, the edges
of the strip approach each other, that is, the wall size decreases.
Thus the wall collapse proceeds also at about the light velocity and
the result of the three-dimensional simulation is confirmed.
In the case of Fig.\ref{fig:obi}, the initial $\dot{\phi}$
is zero everywhere in the box. However, the initial motion may make
the wall move periodically so that the death of wall might be avoided.
We have found that this is not the case by performing the simulations
with various initial motions. Even in the most extreme case where the
strings attached to the wall edges go away each other with the light
velocity, such a motion could do nothing more than the slight
extension of the wall life time.

Finally we have estimated the $\langle E_a \rangle/m_a$ in the
simulation. The time evolution of the potential energy $V_w$ and
kinetic energy of the axion field $A$ is shown in
Fig.\ref{fig:pot}(a). In these simulations, the wall edges are
smoothly connected to the true vacuum region which is different from
the disk wall case. Since the overestimation of the field gradient is
removed by this smoothing, the quantitative analysis of energy
distribution is enabled.  After the wall collapse there remains only
axionic waves which we identify as axions and the values of the
potential and kinetic energy of the axion become constant. Since the
axion waves oscillate like $\sim e^{-iE_at}$, the ratio of the kinetic
energy to potential energy is equal to $\langle E_a \rangle^2/m_a^2$.
The simulation shows $\langle E_a \rangle/m_a \simeq 3$, which means
that the emitted axion is mildly relativistic and becomes
non-relativistic soon by cooling due to the cosmic expansion. As
mentioned before we check the accuracy of our simulation by the
conservation of the total energy in the simulation box. In the
simulation above the total energy is conserved within accuracy of 20\%.

In order to confirm the estimated value of $\langle E_a \rangle/m_a $
we have performed the simulation of higher resolution ( string
width $= 3$ meshes, wall width $= 30$ meshes, wall length $=60$
meshes) . The result (Fig.\ref{fig:pot}(b)) shows $\langle E_a
\rangle/m_a \simeq 3$ which is quite consistent with the result of
the lower resolution run.

Our model parameters in the simulations above correspond to $m_a/F_a =
0.1$ which is much larger than the actual axion model ( $ m_a/F_a \sim
10^{-23}$ ). To check if the estimation for $\langle E_a \rangle/m_a$
depends on $m_a/F_a$, we have run the simulation of thicker wall
( string width $= 1$ meshes wall width $= 50$ meshes, wall length $=100$
meshes ) in which $m_a/F_a =0.02$. Although the conservation of the total
energy becomes as bad as 50\% in this case, the result indicates that
the value of $\langle E_a \rangle/m_a$ is affected only slightly, which
means the generality of our estimation concerning the relativisity of
emitted axions.

\section{Conclusion}
\label{sec:conclusion}

In the previous section, it is found that the axionic wall collapses
with time scale $R/c$ and the mildly-relativistic axions remain
after the collapse. The estimated $\langle E_a \rangle/m_a$ is about 3,
which leads to the relic density of the axion given by
\begin{equation}
    \Omega_a h^2 \simeq  0.01 \alpha^{-3}\beta^2
    \left(\frac{F_a}{10^{10}\mbox{GeV}}\right)^{1.18}
    \label{omega-axion2}
\end{equation}
Since $\alpha$ and $\beta$ are expected to be $O(1)$, the contribution
of the axion to the present universe is large for $F_a \gtrsim
10^{10}\mbox{GeV}$ (compared with the baryon density $\Omega_B h^2 \simeq
0.013$\cite{Walker}) and might account for the dark matter of the
universe. However eq.(\ref{omega-axion2}) should be compared with the
density of the coherent oscillation of axion field (cold
axion)\cite{Preskill,Turner} given by
\begin{equation}
    \Omega_a({\rm cold}) = 6.5\times 10^{-3\pm 0.4}
    \left(\frac{F_a}{10^{10}\mbox{GeV}}\right)^{1.18}.
    \label{cold}
\end{equation}
Therefore the axion from the wall has comparable to or higher density
than the cold axion unless $\alpha^{-3}\beta^2$ is much less than 1.
In deriving (\ref{omega-axion2}), we assume that the domain wall
collapses rapidly after its formation.  As shown in the simulations
this assumption is quite reasonable for walls whose size is much
smaller than the horizon.  However, a large wall does not shrink until
its size becomes smaller than the horizon, which leads to the increase
of the density of axionic domain wall and as a result the density of
the axion produced by the collapse of the wall. Therefore
eq.(\ref{omega-axion2}) may underestimate the actual axion density and
the axion from axionic walls may be more important than the cold axion.

Another important source of axions is an oscillating axionic string.
Let us compare our result with the axion emission from the axionic
strings. The density of the axion from the strings is given by
\begin{equation}
    \Omega_a({\rm string}) \simeq (1 - 0.01) (F_a/10^{10}\mbox{GeV})^{1.18},
    \label{string}
\end{equation}
where uncertainty of a factor of 100 is due to two different
estimations by Davis\cite{Davis} and Harari and Sikivie\cite{Hagmann}.
Thus the importance of the axion from the wall depends on which
estimation is correct.  The difference comes from the different
assumption for the energy spectrum of the emitted axion and it is hard to
judge which assumption is better.  In the case of the axionic domain
wall the spectrum or average energy of emitted axions can be estimated
by numerical simulations more easily since axions are mostly produced
at the final stage of the collapsing wall whose size is much smaller
than the cosmic scale.

In conclusion, the relic density of the axion produced by the collapse
of axionic walls is, at least, larger than the baryon density for $F_a
\gtrsim 10^{10}$ GeV and accounts for a part or all of the dark matter in
the universe.  The axion from the axionic domain wall is more important
than the cold axion and comparable to that from strings if Harari and
Sikivie's estimation is correct.  To make a more precise prediction of
the relic density we have to know the numeric parameter $\alpha$ and
$\beta$, which is beyond the scope of the present work and will be
studied in future work.

\vskip 0.5cm
MN is grateful to Professor K. Sato for his continuous encouragement.
This work was in part supported by the Japanese Grant in Aid for
Science Research Fund of the Ministry of Education, Science and
Culture (No. 3253). Numerical computations were partially performed by
SUN SPARC stations at Uji Research Center, Yukawa Institute for
Theoretical Physics, Kyoto University.

\begin{figure}
\caption{The situation that two walls pass through each other is shown.
The simulation box is one-dimensional and its size is 500 meshes.
The vertical axis depicts the false vacuum energy by $A$ field, $V_w$
which is normalized as $V_w(\pi F_a)=0.02$.
The initial separation is 200 meshes and the initial relative velocity is
$0.05c$. Four figures correspond to the $V_w$ distribution at $t=0$, $t=1500$,
$t=3000$, and $t=5000$ respectively.}
\label{fig:pass}
\end{figure}

\begin{figure}
\caption{Evolution of a strip wall whose length is 20 meshes is shown.
The size of the simulation box is $300^2$. Figures pick out
$100^2$ part of the whole plane. The value of $V_w$ is plotted and
it is normalization as $V_w(\pi F_a)=0.02$.
The first figure is the initial configuration,
the second one is that at $t=10$, and the last is drawn at $t=20$.}
\label{fig:obi}
\end{figure}

\begin{figure}
\caption{Time evolution of the kinetic energy and the potential energy of
the strip walls is shown. The solid lines depict the kinetic energy,
$\frac{1}{2}\dot{A}^2$ and the dashed lines are the part of the potential
energy by the domain wall, $V_w$. a) The same case as Figure (2).
b) string width $= 3$ meshes, wall width $= 30$ meshes, wall length $=60$
meshes.}
\label{fig:pot}
\end{figure}

\end{document}